\documentclass[runningheads]{llncs}
\usepackage{graphicx}
\usepackage{todonotes}
\usepackage{amsmath}
\usepackage{cite}
\usepackage{url}
\usepackage{float}
\renewcommand\UrlFont{\color{blue}\rmfamily}

\begin{document}

\title{Enhancing Energy-efficiency by Solving the Throughput Bottleneck of LSTM Cells for Embedded FPGAs}

\author{Chao Qian, Tianheng Ling, Gregor Schiele}
\date{July 2022}
\institute{University of Duisburg-Essen\\
\email{\{chao.qian, gregor.schiele\}@uni-due.de, tianheng.ling@stud.uni-due.de}\\
}

\authorrunning{C. Qian et al.}
\titlerunning{Energy-efficient LSTM cells for embedded FPGAs}%
\maketitle              
\begin{abstract}

To process sensor data in the Internet of Things(IoTs), embedded deep learning for 1-dimensional data is an important technique. In the past, CNNs were frequently used because they are simple to optimise for special embedded hardware such as FPGAs. This work proposes a novel LSTM cell optimisation aimed at energy-efficient inference on end devices. Using the traffic speed prediction as a case study, a vanilla LSTM model with the optimised LSTM cell achieves 17534 inferences per second while consuming only 3.8 $\mu$J per inference on the FPGA \textit{XC7S15} from \textit{Spartan-7} family. It achieves at least 5.4$\times$ faster throughput and 1.37$\times$ more energy efficient than existing approaches.
\keywords{IoT \and LSTM cell \and Energy-Efficiency \and embedded FPGA}
\end{abstract}

\section{Introduction}
Time-series analysis is a crucial topic in \textit{Machine Learning}. The introduction of the \textit{Long Short-Term Memory} (LSTM) model has significantly enhanced the accuracy of time-series analysis. In IoT-related application scenarios, such as temperature forecast and traffic speed prediction, end devices often send data to the Cloud, where data is analysed using the LSTM model. However, once the Internet connection to the Cloud is broken, end devices can no longer analyse. Additionally, the data transmission consumes energy. For these reasons, on-device intelligence, i.e., performing the LSTM model on the end devices, is preferable.

However, IoT devices' microprocessors (MCUs) are often designed with limited memory and processing power to meet constraint power and energy budgets. Performing LSTM model inference on MCUs is therefore a challenge. Our approach is to add a Field Programmable Gate Array (FPGA) as additional computational power to create energy-efficient LSTM accelerators on it. Our paper's contributions are summarised below:

\begin{itemize}
    \item We propose a novel optimised LSTM cell for embedded FPGAs. It is implemented with VHDL, with a combination of optimisation methods. By solving the throughput bottleneck of the LSTM cell, we maximise its performance while improving the energy efficiency of FPGAs.
    
    \item We apply the optimised LSTM cell in the vanilla LSTM model for traffic speed prediction and verify its performance in real-world applications. The optimised LSTM cell is also applicable to other time series analysis tasks in IoT application scenarios.
    
    \item We integrate the optimised LSTM cell into a PyTorch-based code generation tool, the \emph{elastic-ai.creator}\footnote{\UrlFont{https://github.com/es-ude/elastic-ai.creator}}, enabling developers to easily use our approach to develop their accelerators for FPGAs.
\end{itemize}

We present our work by first discussing some related work in Section \ref{sec:related_work}, followed by background of the LSTM model (especially the LSTM cell in it) in Section \ref{sec:vanilla_lstm_model}. After explaining the optimisations of the LSTM cell in Section \ref{sec:optimise_lstm}, we give an evaluation of this work in Section \ref{sec:evaluation}. Finally, we conclude this paper and plan our future work in Section \ref{sec:conclusion}.

%
\section{Related Work}
\label{sec:related_work}
Previous research applied FPGAs as accelerators to assist LSTM model inference on the Cloud. For example, Cao et al. \cite{cao2019efficient} focused on improving inference speed. However, their approach consumes up to 19 W power and is therefore unsuitable for energy-sensitive IoT application scenarios. With the increased demand for near-end computing, Azari et al. \cite{azari2019energy} proposed a more energy-efficient FPGA-based LSTM accelerator. Although they optimised the power consumption of the FPGA to 1.19 W, their approach is still too expensive for long-term monitoring with battery power.

Recently, researchers have started considering the energy efficiency of on-device FPGA accelerators. In 2020, Hasib-Al-Rashid et al. \cite{manjunath2020low} proposed a LSTM processor for FPGA \textit{XC7A100T} from \textit{Artix-7} family. Their design only utilises 1\% of LUTs, 9\% of BRAM and 1.67\% of DSP slices of this FPGA by extremely reusing the hardware resources, such as only implementing two multiplication and accumulation (MAC) units in the LSTM cell. It only consumes 17 mW dynamic power and performs 0.055 GOP/s. However, idling the hardware resources of the FPGA does not reduce the static power consumption of the FPGA, which was estimated as 92 mW. The high portion of static power leads to poor overall energy efficiency (0.5 GOP/J). 

Noticing this problem, Chen et al. \cite{chen2021eciton} implemented a similar accelerator on a much smaller FPGA \textit{iCE40 UP5K} in 2021. Thanks to the ultra-low static power (at $\mu W$ scale) of this FPGA, the overall power consumption during inference is approximately equal to the dynamic power of the FPGA, which is 17 mW. The energy efficiency is increased to 3.9 GOP/J, while the throughput is slightly improved to 0.067 GOP/s due to the same parallelism strategy they applied as in \cite{manjunath2020low}.

In both works \cite{manjunath2020low, chen2021eciton}, researchers applied fixed-point logic to simplify the design and reduce the loss of precision compared to aggressive quantisation. The activation functions $tanh()$ and $sigmoid()$ were replaced with $hard\_tanh()$ and $hard\_sigmoid()$ respectively, which simplifies the computations but leads to a large reduction of precision \cite{otte2014dynamic}. Namin et al. and Meher et al. used lookup tables to implement activation functions for higher speed and precision \cite{namin2009artificial, meher2010optimized}.

Inspired by this previous work, we chose a slightly larger FPGA than the \textit{iCE40 UP5K}, the \textit{XC7S15}, which has about 2.5 times LUTs, 10 times BRAM and 2.5 times DSP. Therefore, it can afford higher parallelism and deeper lookup tables with wider fixed-point data for higher precision. As it has around 10\% of resources of the \textit{XC7A100T}, it still consumes relatively low static power. 

\section{LSTM Background and Analysis}
\label{sec:vanilla_lstm_model}
Before presenting our optimisations, we first introduce the basic concepts of LSTM models, layers and cells. We then analyse the timing complexity of a single LSTM cell to determine the optimisation potential. 

\subsection{LSTM Model vs Layer vs Cell}

\begin{figure}[!htb]
    \centering
    \includegraphics[width=\textwidth]{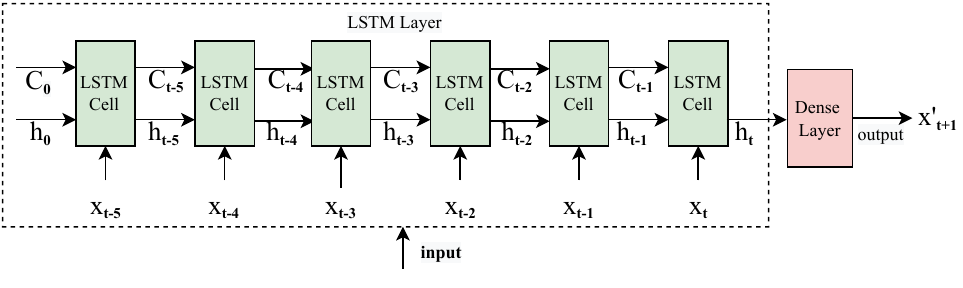}
    \caption{The unfolded architecture of the LSTM model in the time dimension}
    \label{fig:Vanilla_LSTM}
\end{figure}

The most conventional and simplistic LSTM model is constructed with one LSTM layer followed by a dense layer. By setting an activation function in the dense layer, the LSTM model can perform regression or classification tasks. We use an LSTM model (see Figure \ref{fig:Vanilla_LSTM}) taken from \cite{fu2016using}, to predict traffic speed. It takes 6 historical data points as inputs ($x_{t-5}, ..., x_{t-1}, x_t$), and predicts the next data point (${x'}_{t+1}$) as its output. 

Inside the LSTM layer, a single LSTM cell processes the 6 input data points recurrently to perform the long and short-term memory logic, which is visualized as 6 recurrent steps in the time dimension in Figure \ref{fig:Vanilla_LSTM}. 

\begin{figure}[!htbp]
\centering
\includegraphics[width=\textwidth]{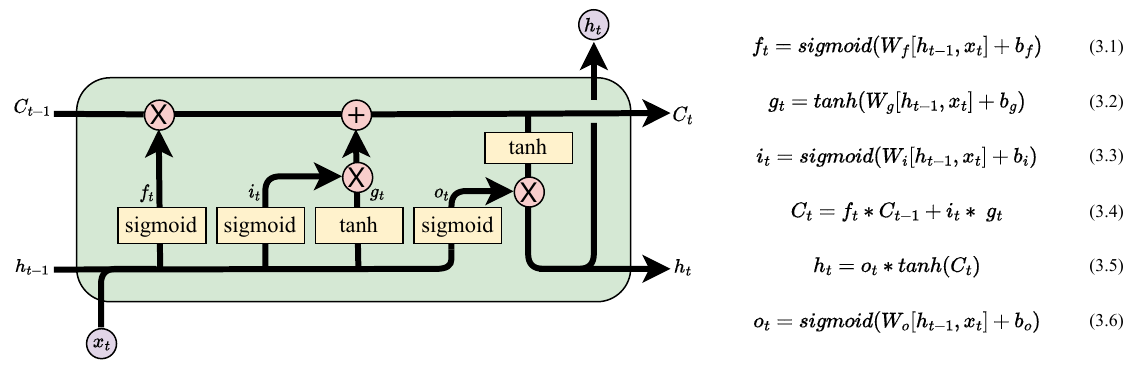}
\caption{The structure of an LSTM cell}
\label{fig:lstm_cell}
\end{figure}

In the LSTM cell (see Figure \ref{fig:lstm_cell}) the information is carried through the sequence chain in the cell state $C_t$ and the hidden state $h_{t}$. Internally, the cell contains three so-called gates, the input gate $i_t$, the output gate $o_t$ and the forget gate $f_t$ to control which information to retain or discard. All computations that happen in the LSTM cell can be described by Equations 3.1 to 3.6, which are explained in detail in \cite{lstm_cell}. We use \(\ast\) to denote the Hadamard product, \([\cdot, \cdot]\) to denote the concatenation of two vectors. Weight matrices are denoted by \(W\). 

An important factor affecting the model performance is the size of the hidden state, i.e. how many neurons are necessary to represent it. Commonly used hidden sizes include 1 \cite{fu2016using}, 12 \cite{manjunath2020low}, 32 to 256 \cite{zhang2017power}. In our application case we use a hidden size of 20. The resulting model has a high accuracy on the test set, while being small enough to fit our target FPGA.

\subsection{Timing}

\begin{figure}[!ht]
    \centering
    \includegraphics[width=0.9\textwidth]{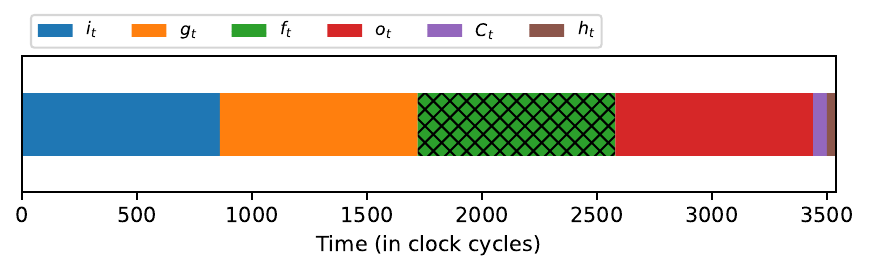}
    \caption{Time breakdown of a single recursion in a sequentially processed LSTM cell} 
    \label{fig:timing_paralell}
\end{figure}

As Figure \ref{fig:Vanilla_LSTM} shows, the computation of each recursion depends on the result of the previous recursion, so increasing the number of LSTM cells in the LSTM layer cannot help to improve throughput. A possible way to increase the throughput of the LSTM model is to reduce the time spent processing an LSTM cell. Based on the timing model of the sequentially executed LSTM model in \cite{sun2018fpga}, we plotted the timing decomposition of a sequentially processed LSTM cell whose input\_size is 1 and hidden\_size is 20. 
Figure \ref{fig:timing_paralell} shows that the processing of Equation 3.1, 3.2, 3.3 and 3.6 take up 97.1\% of the time to process the whole cell, indicating that the throughput bottleneck of the LSTM cell lies in the computations of these four equations. In contrast, the operations in the dense layer only consume 0.6\% of the time. 

\section{Optimised LSTM Cell Design}
\label{sec:optimise_lstm}

This section begins with a description of the parallelisation optimisation in the LSTM cell. Then, we will discuss memory optimisation.

\subsection{Parallelising the LSTM Cell}

\begin{figure}[!htb]
    \centering
    \includegraphics[width=0.9\textwidth]{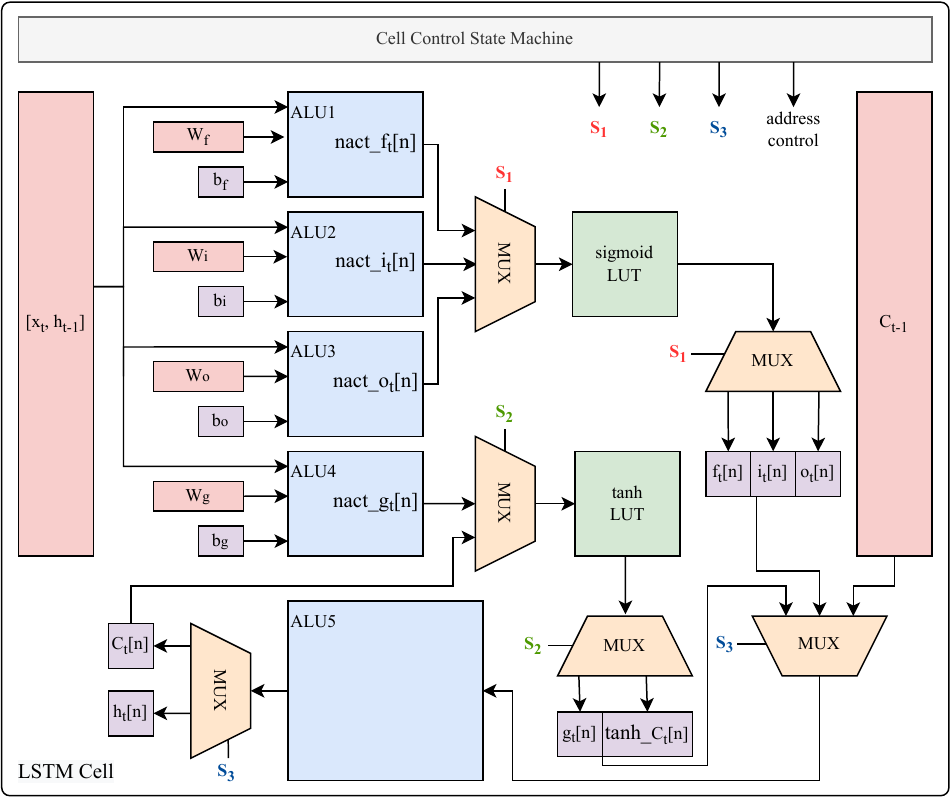}
    \caption{Architecture of our LSTM cell on an FPGA. $nact\_f_t$ denotes that $f_t$ is not activated, others have similar meaning.} 
    \label{fig:lstm_archietecture}
\end{figure}

To solve the mentioned performance bottleneck, we analysed the dependencies of Equations  3.1, 3.2, 3.3 and 3.6. They take the same data ($x_t$, $h_{t-1}$) and process it independently, i.e., they can be computed simultaneously. We therefore create four identical ALU modules (see \textit{ALU1}, \textit{ALU2}, \textit{ALU3}, and \textit{ALU4} in Figure \ref{fig:lstm_archietecture}). Each of them consumes 1 DSP slice. These ALU modules are used to execute multiply-accumulate operations of these equations concurrently. By quadrupling the number of ALU modules, the computation speed of gates in the LSTM cell is increased by a factor of four, making the increase in throughput possible. Note that previous work used two DSP slides, one for calculating $W_i x_t$, another one for $W_h h_{t-1}$\cite{manjunath2020low,chen2021eciton}. Due to the different complexity of $x_t$ and $h_{t-1}$, this leads to bad utilisation of the DSPs. Our assignment avoids this problem.

Our ALU modules do not include activation functions. Instead, we instantiate a lookup table for each kind of activation function (see the \textit{sigmoid LUT} module and the \textit{tanh LUT} module in Figure \ref{fig:lstm_archietecture}). It is well known that the greater the depth of the lookup table, the smaller the degradation in the accuracy of the model inference. By placing the lookup tables once and sharing them on demand, the optimised LSTM cell can save more hardware resources, which can be used to implement larger lookup tables, helping to improve the precision of the activation function. However, even with four ALU modules running concurrently, waiting for the whole matrix multiplication (e.g., $W_f [h_{t-1}, x_t]$) to finish before updating $C_t$ and $h_t$ still takes unacceptable time. Thus, the longest stage of our pipeline is only for computing one row of the matrix multiplication. Once new elements with index \textit{n} ($f_t[n]$, $i_t[n]$, $g_t[n]$, $o_t[n]$) are computed, the computation for $C_t[n]$, $h_t[n]$ can start.

As shown in Figure \ref{fig:lstm_archietecture}, we add another ALU module (\textit{ALU5}) to execute the multiply-accumulate operations in Equations 3.4 and 3.5. Since these operations are less complex than the others, it is possible to reuse a single ALU module for both of them without reducing the overall performance. However, this is only true for cells with a larger hidden size. Since we had free DSPs available, we decided to implement \textit{ALU5} with three DSPs. This makes our design suitable for cells with smaller hidden sizes (down to 3). 

\begin{figure}[!htb]
    \centering
    \includegraphics[width=0.9\textwidth]{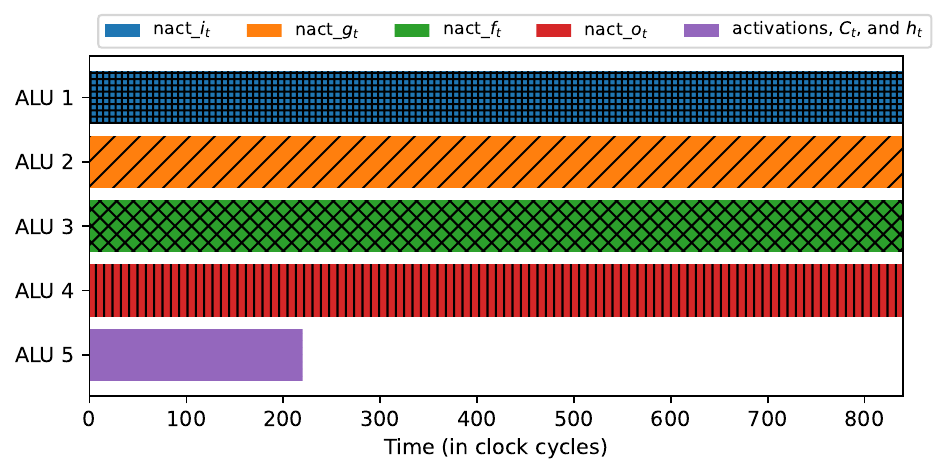}
    \caption{Time breakdown of a single recursion in a parallelly processed LSTM cell} 
    \label{fig:timing_sequential}
\end{figure}

Figure \ref{fig:timing_sequential} shows the time breakdown when the above-mentioned parallel computing is applied in an LSTM cell. Although the processing time of each ALU for its corresponding equation has not improved, the overall time consumption of the entire recursion is squeezed to 860 clock cycles, which would give us a 4.1-fold improvement in throughput compared to sequential computing.

\subsection{Memory Management}

Our approach uses only on-chip memory, so-called Block- and Distributed-RAM (BRAM and LUTRAM), to store data (including parameters and intermediate results) on the FPGA. On-chip memory is relatively limited in capacity but extremely fast and flexible in terms of placement, construction and connection.

Since the inputs required for the computation performed by \textit{ALU1} to \textit{ALU4} are the same, we store $x_t$ and $h_{t-1}$ in the same memory module. Due to the flexible wiring feature of the FPGA, these four ALU modules share the data in the memory and simultaneously share the read-out data via a data bus.

Furthermore, the static parameters (i.e., weights and bias) are stored in the bitstream and automatically initialised at startup-time. Therefore, the overhead of loading parameters is eliminated at run-time. Eliminating circuits for loading parameters can significantly simplify the design further.
In addition, weights and bias are separately allocated and placed close to their corresponding ALU to minimise the signal delay.

\section{Evaluation}
\label{sec:evaluation}

To discuss our evaluation, we first describe the used data set and our model implementation. Then we present our evaluation results with respect to FPGA resource utilisation, timing, and power consumption. Finally, we compare our results to other approaches. 

\subsection{Data Set and Training Settings}

For our experiments, we aimed to predict traffic speed. We used the publicly available data set \textbf{PeMS-4W}\footnote{\UrlFont{https://doi.org/10.5281/zenodo.3939793}}, which contains 11,160 time series corresponding to sensor measurements at different locations of sensors over four weeks. Each time series contains a measurement every 5 minutes, leading to 8064 time points. From these, we randomly selected one time series to form our data set. We divided it into a training set and a test set in a ratio of 3:1.
Our full-precision (double-precision floating-point) LSTM model was trained on the selected training set with 30 epochs with a batch size of 1. We used an \textit{Adam} optimiser with $beta_1=0.9$, $beta_2=0.98$ and $epsilon=10^{-9}$. The initial learning rate was set to 0.01, while a learning rate scheduler with $step\_size=3$ and $gamma=0.5$ was used. Mean-Squared Loss was used as the loss function, and Mean Squared Error (MSE) as the evaluation metric. The MSE of the trained full precision LSTM model is 0.1663.

\subsection{Model Implementation}
\label{subsec:eval_fixed_point_quant}

We implemented our optimised LSTM cell using our \textit{elastic-ai.creator} tool. Given a trained PyTorch model, the \textit{elastic-ai.creator} can automatically transform a full precision model into an optimised model and translate it into VHDL code for an FPGA. The \textit{elastic-ai.creator} includes implementations for different layers. Most importantly for us, it contains a built-in optimised dense layer that uses only 1 DSP slice (see \cite{burger2020embedded}). We used this for our model and extended the tool with an implementation of our LSTM cell design.

We performed post-training quantisation to fixed-point representations on the LSTM model. We describe this representation with a notation of ($x$, $y$), where $x$ is the number of fractional bits (representing numbers < 1) and $y$ is the total width in bits \cite{burger2020embedded}. 

\begin{figure}
    \centering
    \includegraphics[width=0.65\textwidth]{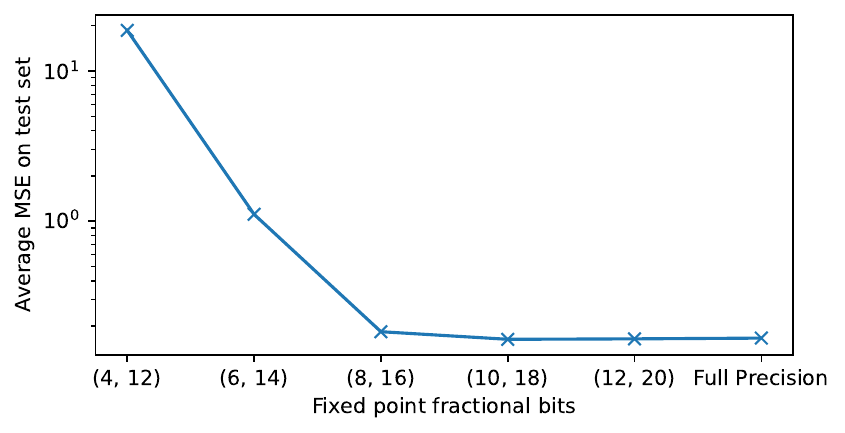}
    \caption{The MSE of the quantised LSTM model on test set with various fractional bits}
    \label{fig:mse_with_diffbits}
\end{figure}

To evaluate the effect of the fractional digit $x$ on the model inference, we varied it from 4 to 12 while utilizing 8 bits for the integer part. We performed this on a custom Python simulator with all parameters and variables at the corresponding fixed-point width. We kept the activation function in full precision. Figure \ref{fig:mse_with_diffbits} shows that the MSE (0.1722) on the test set no longer drops significantly when $x$ is greater than 8. Therefore, our experiments below are conducted with a fixed-point configuration of (8, 16). Clearly, this can be optimised further in the future. Our design supports scalable bit-width for fixed-point data so that AI developers can choose other settings based on the output of the Python simulator. 

Furthermore, we replaced all full precision activation functions with lookup tables of depth 256. We conducted experiments with different lookup table depths both in our simulator and on an \textit{XC7S6} FPGA. The results differ slightly, which may be due to their different rounding mechanisms.
As shown in Table \ref{tab:lut_depth_vs_mes}, the MSE on the test set decreases as the depth of the lookup table increases. Note that the depth of the lookup tables is the same for different activation functions. At depth 256, the MSE (0.1821) is close enough to the MSE with full precision activation functions (0.1722). Since we instantiate only one lookup table for each type of activation function and share it over time, even such lookup tables with larger depths can be used on embedded FPGAs.

\begin{table}[htb]
\centering
\caption{The MSE of the test set when using lookup tables of different depths}
\begin{tabular}{|c|c|c|}
\hline
\textbf{  Depth of lookup table  } & \textbf{  MSE on Python Simulator  } & \textbf{  MSE on FPGA }   \\ \hline
\textbf{64} & 0.6920 & 0.6833 \\ \hline
\textbf{128} & 0.2485 & 0.2491 \\ \hline
\textbf{256} & 0.1821 & 0.1659 \\ \hline
\end{tabular}
\label{tab:lut_depth_vs_mes}
\end{table}

\subsection{Resource utilisation on FPGA}

Using a fixed-point configuration (8,16) and lookup tables of depth 256, we analysed the resource consumption using a synthesis tool integrated into the \textit{Vivado} IDE from \textit{Xilinx}. We synthesised our design into three target FPGAs of the \textit{Spartan-7} family with different on-chip resources. 

As shown in Table \ref{tab:resource_overview}, our design can fit the \textit{XC7S6} which is the smallest FPGA of the \textit{Spartan-7} family. Although 80\% of the DSP slices (the most critical resource) are used, we still have a considerable amount of other available resources to make the design more accurate. Our optimised LSTM cell requires seven DSP slices, while the dense layer consumes one DSP slice, so we could, e.g., add two more dense layers to the model and still fit onto the \textit{XC7S6}. 

Furthermore, the resource utilisation on the \textit{XC7S15} is below 50\% for all types, which means that we can either deploy two such LSTM models on this device or at least double the number of layers of the current LSTM model. The \textit{XC7S25} FPGA has 1.8 times more LUTs and 4 times more DSPs than the \textit{XC7S15}, so there is no doubt that the \textit{XC7S25} can be the choice for more complex models using our optimisation method.

\begin{table}[htb]
\centering
\caption{Resource utilisation on Spartan-7 FPGAs}
\begin{tabular}{|l|r|r|r|c|}
\hline
 & \multicolumn{1}{|c|}{\textbf{}}            & \multicolumn{3}{|c|}{\textbf{Utilisation on FPGAs (in \%)}}                                                \\ \cline{3-5} 
\multicolumn{1}{|c|}{\textbf{Resource}}                & \multicolumn{1}{|c|}{\textbf{Estimation}} & \multicolumn{1}{|c|}{\textbf{XC7S6}} & \multicolumn{1}{|c|}{\textbf{XC7S15}} & \multicolumn{1}{|c|}{\textbf{XC7S25}} \\ \hline
\multicolumn{1}{|c|}{\textbf{LUT}}    & \multicolumn{1}{|c|}{1435}                & \multicolumn{1}{|c|}{38.3}        & \multicolumn{1}{|c|}{17.9}         & 9.8                              \\
\hline
\multicolumn{1}{|c|}{\textbf{LUTRAM}} & \multicolumn{1}{|c|}{60}                  & \multicolumn{1}{|c|}{2.5}         & \multicolumn{1}{|c|}{2.5}          & 1.2                              \\
\hline
\multicolumn{1}{|c|}{\textbf{BRAM}}   & \multicolumn{1}{|c|}{2}                   & \multicolumn{1}{|c|}{40.0}          & \multicolumn{1}{|c|}{20.0}           & 4.4                              \\
\hline
\multicolumn{1}{|c|}{\textbf{DSP}}    & \multicolumn{1}{|c|}{8}                   & \multicolumn{1}{|c|}{80.0}          & \multicolumn{1}{|c|}{40.0}           & 10.0    \\
\hline
\end{tabular}
\label{tab:resource_overview}
\end{table}

\subsection{Processing Time}
Our design does not follow the usual CPU-based sequential computing nor a large FPGA-based parallel computing. Therefore, we proposed a timing model to estimate the system throughput.

By calculating the required number of operations $n_{total}$, we can estimate the processing time of the LSTM model. The simplified timing model we have designed is defined by Equation \ref{equ:t_model}, where $t_{clock}$ is the reference clock period of the FPGA, $n_{ll}$ and $n_{dense}$ are the clock cycles of the LSTM layer and the dense layer respectively. 

\begin{figure}[!htb]
    \begin{equation}
    t_{model} = t_{clock} * n_{total}= t_{clock} * (n_{ll}+n_{dense})
    \tag{5.1}
    \label{equ:t_model}
    \end{equation}
    \begin{equation}
    n_{ll} = n_{seq} * n_{lc} =  n_{step} * (n_i+n_h) * 2 * (n_h+1)
    \tag{5.2}
    \label{equ:n_ll}
    \end{equation}
    \begin{equation}
    n_{dense} = n_f * n_o * 2
    \tag{5.3}
    \label{equ:n_dense}
    \end{equation}
\end{figure}

$n_{ll}$ can be further represented by Equation \ref{equ:n_ll}, where $n_{seq}$ represents the length of input sequences, $n_i$ and $n_h$ are the \textit{input\_size} and \textit{hidden\_size} of the LSTM cell respectively. Furthermore, the factor \textit{2} indicates that our ALU module requires 2 clock cycles to produce an output. Similarly, $n_{dense}$ can be defined by Equation \ref{equ:n_dense}, where $n_f$ and $n_o$ represent the number of input and output features of the dense layer. For our model structure, $n_f$ is always equal to $n_h$ since only the latest hidden state is fed into the dense layer.

These equations allow the processing time of our LSTM model to be estimated. The total number of clock cycles $n_{total}$ is 5332. Assuming that we deploy the LSTM model on the \textit{XC7S15} with a clock frequency of 100 MHz, the estimated processing time is 53.32 $\mu$s. Our design can then process up to 18754 samples per second.

We validated the timing of the LSTM model on the actual \textit{XC7S15}. The processing time measured in hardware with a 100 MHz clock is 57.25 $\mu$s, which is 3.93 $\mu$s more than the estimated processing time. Although there is still some discrepancy, it proves that our timing model is valid.

\subsection{Inference Power}
Using the XPE software, we estimated the power consumption during the inference of the LSTM model for the target FPGAs. We also applied the calibration process according to the guidelines provided by \textit{Xilinx} \cite{Xilinx:ug997} to improve the confidence of the estimation. The estimated power can be divided into static and dynamic power.

\begin{figure}
    \centering
    \includegraphics[width=0.6\textwidth]{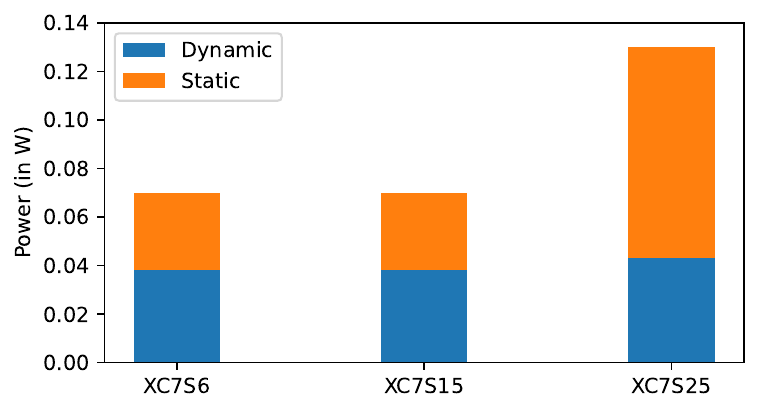}
    \caption{Power estimation for inference on different FPGAs}
    \label{fig:power_estimation}
\end{figure}

The static power consumption of the FPGAs is characterised by the chip design and independent of switching activity. As Figure \ref{fig:power_estimation} shows, the static power of the \textit{XC7S6} and \textit{XC7S15} is identical (32 mW), while the static power of the \textit{XC7S25} is much higher (87 mW). We infer that the \textit{XC7S6} and \textit{XC7S15} use the same chip (with some resources on the \textit{XC7S6} not being available to users). 

The dynamic power is modelled based on the switching activity during inference. 
The dynamic power of both \textit{XC7S6} and \textit{XC7S15} is 38 mW. In comparison, the dynamic power of the \textit{XC7S25} is 43 mW. 

Based on the estimated power consumption and the processing time on the \textit{XC7S15}, the estimated energy per inference is 3.7 $\mu$J. On the actual \textit{XC7S15} hardware, the measured energy per inference is 4.1 $\mu$J.

\subsection{Comparison with the State-of-the-Art}

As a final evaluation step, we computed the throughput and energy efficiency of our LSTM model on real hardware (see Table \ref{tab:comparison}). Our model is 5.4 times faster than \cite{chen2021eciton} and 6.6 times faster than \cite{manjunath2020low}. The energy efficiency of our model is 10.66 times higher than \cite{manjunath2020low}. 
Although the \textit{XC7S15} in our work consumes more static power than the \textit{iCE40 UP5K} in \cite{chen2021eciton}, due to the improved throughput, the energy efficiency of our model is still 1.37 times higher than theirs.

\begin{table}[htb]
\centering
\caption{Compare throughput and energy efficiency with the state-of-the-art}
\begin{tabular}{|l|c|c|c|c|}
\hline
                            & This Work        & \cite{chen2021eciton}  & \cite{manjunath2020low}  \\ \hline
Platform                    & \textit{XC7S15}           & \textit{iCE40 UP5K}             &        \textit{XC7A100T} \\ \hline
Clock (MHz)                 &  \textbf{100}         &  17                    &        52.6          \\ \hline
Power (mW)                  &  \textbf{71}           &  17                    &       109          \\ \hline 
Throughput (GOP/s)         & \textbf{0.363}         &  0.067                 &       0.055        \\ \hline
Energy efficiency (GOP/J) & \textbf{5.33}          &  3.9                   &       0.5          \\ \hline 
\end{tabular}
\label{tab:comparison} 
\end{table}

\section{Conclusion and Outlook}
\label{sec:conclusion}
Energy-efficient artificial intelligence on end devices enables interesting IoT applications. It offers the opportunity to ensure the quality of IoT services without relying on the Internet connection.

Our approach improves the energy efficiency of the LSTM cell by solving its throughput bottleneck. The model with optimised LSTM cell achieves 17534 inferences per second with only 71 mW power consumption. Its super high energy efficiency (5.33 GOP/J) can promise longer battery life for continuous analysis of time series data on the device.

In the future, we plan to increase throughput by achieving lower bit quantisation through quantisation-aware training. In addition, we will validate our approach on further time-series classification tasks, enabling users to solve more targeted applications.

\subsection*{Acknowledgements}
The authors acknowledge the financial support by the Federal Ministry of Education and Research of Germany in the KI-LiveS project.

\bibliographystyle{splncs04}
\bibliography{reference}

\end{document}